\documentclass[11pt,a4paper]{article}

\usepackage[utf8]{inputenc}
\usepackage[T1]{fontenc}
\usepackage{lmodern}
\usepackage[margin=2.5cm]{geometry}
\usepackage{amsmath,amssymb}
\usepackage{graphicx}
\usepackage{booktabs}
\usepackage{hyperref}
\usepackage{xcolor}
\usepackage{natbib}
\usepackage{caption}
\usepackage{subcaption}
\usepackage{enumitem}
\usepackage{pgfplots}
\pgfplotsset{compat=1.18}
\usepackage{tikz}
\usetikzlibrary{patterns}

\hypersetup{
    colorlinks=true,
    linkcolor=blue!60!black,
    citecolor=blue!60!black,
    urlcolor=blue!60!black,
}

\title{Fine-tuning RoBERTa for CVE-to-CWE Classification:\\A 125M Parameter Model Competitive with LLMs}

\author{Nikita Mosievskiy\\
Independent Researcher\\
\texttt{https://huggingface.co/xamxte}}

\date{}

\begin{document}

\maketitle

\begin{abstract}
We present a fine-tuned RoBERTa-base classifier (125M parameters) for mapping Common Vulnerabilities and Exposures (CVE) descriptions to Common Weakness Enumeration (CWE) categories. We construct a large-scale training dataset of 234,770 CVE descriptions with AI-refined CWE labels using Claude Sonnet 4.6, and agreement-filtered evaluation sets where NVD and AI labels agree. On our held-out test set (27,780 samples, 205 CWE classes), the model achieves 87.4\% top-1 accuracy and 60.7\% Macro F1---a +15.5 percentage-point Macro F1 gain over a TF-IDF baseline that already reaches 84.9\% top-1, demonstrating the model's advantage on rare weakness categories. On the external CTI-Bench benchmark (NeurIPS 2024), the model achieves 75.6\% strict accuracy (95\% CI: 72.8--78.2\%)---statistically indistinguishable from Cisco Foundation-Sec-8B-Reasoning (75.3\%, 8B parameters) at 64$\times$ fewer parameters. We release the dataset, model, and training code.

\medskip
\noindent\textbf{Resources:}
Dataset: \url{https://huggingface.co/datasets/xamxte/cve-to-cwe} \quad
Model: \url{https://huggingface.co/xamxte/cwe-classifier-roberta-base}
\end{abstract}

\section{Introduction}

Mapping CVE entries to CWE categories is essential for vulnerability triage, prioritization, and remediation. The National Vulnerability Database (NVD) provides CWE labels for most CVEs, but these labels are known to be noisy---often too generic (e.g., CWE-20 ``Improper Input Validation'' used as a catch-all) or inconsistent across annotators.

Recent work has evaluated large language models on this task. The CTI-Bench benchmark \citep{alam2024ctibench} showed that GPT-4 achieves 72.0\% accuracy on CVE-to-CWE mapping, with smaller models performing significantly worse (LLaMA3-8B: 44.7\%). Google's Sec-Gemini~v1 reports outperforming other models by at least 10.5\% on CTI-RCM \citep{google2025secgemini}, but is a closed proprietary system with no exact accuracy published. Cisco's Foundation-Sec-8B-Reasoning (8B parameters) reaches 75.3\% \citep{cisco2025reasoning}.

We show that a much smaller model---RoBERTa-base with 125M parameters---can achieve competitive results through careful dataset construction and two-phase fine-tuning. Our contributions:

\begin{enumerate}[nosep]
    \item A large-scale CVE-to-CWE dataset (290K samples) with AI-refined labels and MITRE ATT\&CK annotations, released publicly;
    \item A two-phase fine-tuning approach that achieves 87.4\% top-1 accuracy on an agreement-filtered test set;
    \item Competitive open-weight results on CTI-Bench RCM with a task-specific encoder 64$\times$ smaller than the nearest competing general-purpose model;
    \item Analysis of CWE hierarchy granularity as a confound in current benchmarks.
\end{enumerate}

\section{Dataset Construction}

\subsection{Source Data}

We collected 336,777 CVE entries (1999--2026) from the National Vulnerability Database. After removing duplicates and entries with insufficient descriptions, 318,979 CVEs remained for processing. Of these, 244,866 had NVD-assigned CWE labels while 74,113 were unlabeled in NVD.

\subsection{Label Refinement with Claude Sonnet 4.6}

NVD CWE labels are assigned by different CNAs (CVE Numbering Authorities) with varying quality. To improve label quality, we relabeled all 318,979 descriptions using Claude Sonnet 4.6 via the Anthropic Batch API ($\sim$\$395 total cost). Each CVE description was independently classified into one of 205 CWE categories.

Among the 244,866 CVEs with existing NVD labels, 73.1\% had exact CWE ID match with the Sonnet prediction. When allowing parent$\leftrightarrow$child CWE equivalences, agreement rises to 84.5\%, indicating that roughly half of the disagreements are granularity differences rather than genuine labeling errors. The remaining 15.5\% disagreement cases were predominantly:
\begin{itemize}[nosep]
    \item NVD using generic parent CWEs where Sonnet assigned specific children (e.g., CWE-119 $\rightarrow$ CWE-121);
    \item Genuinely ambiguous cases with multiple plausible CWEs;
    \item NVD labeling errors (e.g., CWE-79 XSS for what is clearly a SQL injection).
\end{itemize}

\subsection{Split Construction}

We constructed evaluation-optimized splits:
\begin{itemize}[nosep]
    \item \textbf{Validation} (27,896) and \textbf{Test} (27,780): Only samples where NVD and Sonnet labels agree. This \emph{agreement filtering} yields a high-confidence subset, but biases evaluation toward unambiguous cases---samples where labelers disagree are excluded, so the test set is systematically easier than a random sample of NVD.
    \item \textbf{Training} (234,770): All remaining samples using Sonnet labels, including NVD-Sonnet agreement samples, disagreement samples (using Sonnet's label), and 30,100 previously unlabeled CVEs that received Sonnet labels matching one of the 205 target classes.
\end{itemize}

\subsection{MITRE ATT\&CK Mapping}

In addition to CWE labels, we mapped each CVE to MITRE ATT\&CK techniques using Claude Sonnet 4.6, yielding 361 unique technique IDs with 97.2\% coverage. This multi-label annotation is included in the released dataset but is not evaluated in this paper; we include it as a resource for future work on multi-task vulnerability analysis.

\subsection{Data Decontamination}

All 2,000 CVEs from the CTI-Bench benchmark were removed from all splits to ensure clean external evaluation.

\section{Method}

\subsection{Model}

We fine-tune RoBERTa-base \citep{liu2019roberta} with a classification head for 205 CWE classes. Input format: \texttt{"CVE Description: \{text\}"}, max length 384 tokens.

\subsection{Two-Phase Training}

We employ a two-phase training strategy:

\textbf{Phase~1---Classifier warm-up} (4 epochs): Freeze the first 8 of 12 transformer layers and the embedding layer. Only the top 4 layers and the classification head are trained. Learning rate: $10^{-4}$ with cosine schedule.

\textbf{Phase~2---Full fine-tuning} (9 epochs): Unfreeze all parameters. Learning rate: $2 \times 10^{-5}$ with cosine schedule and 6\% warmup.

This approach prevents catastrophic forgetting of pretrained representations while allowing the classifier head to reach a good initialization before full fine-tuning begins.

\subsection{Training Details}

\begin{table}[h]
\centering
\caption{Training hyperparameters.}
\label{tab:hyperparams}
\begin{tabular}{ll}
\toprule
Hyperparameter & Value \\
\midrule
Batch size & 32 \\
Max sequence length & 384 \\
Label smoothing & 0.1 \\
Weight decay & 0.01 \\
Max gradient norm & 1.0 \\
Precision & bf16 \\
Early stopping patience & 10 (phase 2) \\
Hardware & NVIDIA RTX 5080 (16\,GB) \\
Training time & $\sim$4 hours (both phases, wall clock) \\
\bottomrule
\end{tabular}
\end{table}

\subsection{Baseline}

We compare against a TF-IDF baseline (50K features, unigrams + bigrams) with Logistic Regression trained on the same data.

\section{Results}

\subsection{Internal Evaluation}

Results on the held-out test set (27,780 agreement-filtered samples):

\begin{table}[h]
\centering
\caption{Test set results (27,780 agreement-filtered samples). Note that agreement filtering biases toward unambiguous cases; real-world accuracy on arbitrary NVD entries will be lower.}
\label{tab:internal}
\begin{tabular}{lcccc}
\toprule
Model & Top-1 Acc & Top-3 Acc & Macro F1 & Weighted F1 \\
\midrule
\textbf{RoBERTa-base (ours)} & \textbf{87.4\%} & \textbf{94.7\%} & \textbf{0.607} & \textbf{0.872} \\
TF-IDF + Logistic Regression & 84.9\% & 92.7\% & 0.452 & --- \\
\bottomrule
\end{tabular}
\end{table}

The TF-IDF baseline is surprisingly strong on top-1 accuracy (84.9\%), as common CWEs like XSS and SQL injection are easily identified by keyword patterns. However, the gap widens significantly on Macro F1 (+15.5pp), indicating that RoBERTa provides substantially better performance on rare CWE classes.

\textbf{Macro F1 breakdown.}\quad The overall Macro F1 of 0.607 masks significant variation across CWE classes. Of the 205 classes: 64 classes achieve F1 $\geq$ 0.8 (covering $\sim$22K test samples---the well-represented, distinctive CWEs); 104 classes fall in 0.3 $\leq$ F1 $<$ 0.8 ($\sim$5.5K samples); and 32 classes have F1 $<$ 0.3 ($\sim$278 samples---rare CWEs with very few test examples). The low-F1 tail consists almost entirely of classes with $<$10 test samples, where even a single misclassification causes F1 to collapse. For a vulnerability triage system, this means the model is reliable on common weakness types but should be treated with caution on rare CWEs.

\subsection{CTI-Bench External Benchmark}

\begin{figure}[h]
\centering
\begin{tikzpicture}
\begin{axis}[
    xbar,
    width=12cm,
    height=13cm,
    bar width=6pt,
    enlarge y limits={abs=0.6},
    xlabel={CTI-Bench RCM Top-1 Accuracy (\%)},
    xmin=35,
    xmax=92,
    ytick={0,1,2,3,4,5,6,7,8,9,10,11,12},
    yticklabels={
        LLaMA3-8B (8B),
        LLaMA3-70B (70B),
        Gemini 1.5,
        GPT-3.5 (~175B),
        Primus-Base (8B),
        F-Sec-8B-Inst (8B),
        WRabbitNeo-70B (70B),
        F-Sec-8B (8B),
        GPT-4 (~1.7T),
        F-Sec-8B-R (8B),
        \textbf{Ours (125M)},
        SecLM*,
        Sec-Gemini v1*,
    },
    y dir=reverse,
    bar width=8pt,
    nodes near coords,
    nodes near coords align={horizontal},
    every node near coord/.append style={font=\scriptsize},
    tick label style={font=\small},
    label style={font=\small},
    legend style={at={(0.97,0.03)}, anchor=south east, font=\small},
]
\addplot[fill=gray!30, draw=gray!60, bar shift=0pt] coordinates {
    (44.7, 0)
    (65.9, 1)
    (66.6, 2)
    (67.2, 3)
    (67.8, 4)
    (70.4, 5)
    (71.1, 6)
    (72.0, 7)
    (72.0, 8)
    (75.3, 9)
};
\addplot[fill=blue!50, draw=blue!70, bar shift=0pt] coordinates {
    (75.6, 10)
};
\addplot[pattern=north east lines, pattern color=red!40, draw=red!40, bar shift=0pt] coordinates {
    (85.0, 11)
    (86.0, 12)
};
\end{axis}
\end{tikzpicture}
\caption{CTI-Bench RCM comparison. Our task-specific 125M-parameter model (blue) is competitive with general-purpose 8B-parameter models (gray). Hatched bars indicate Google's closed systems with approximate scores estimated from published charts.}
\label{fig:comparison}
\end{figure}
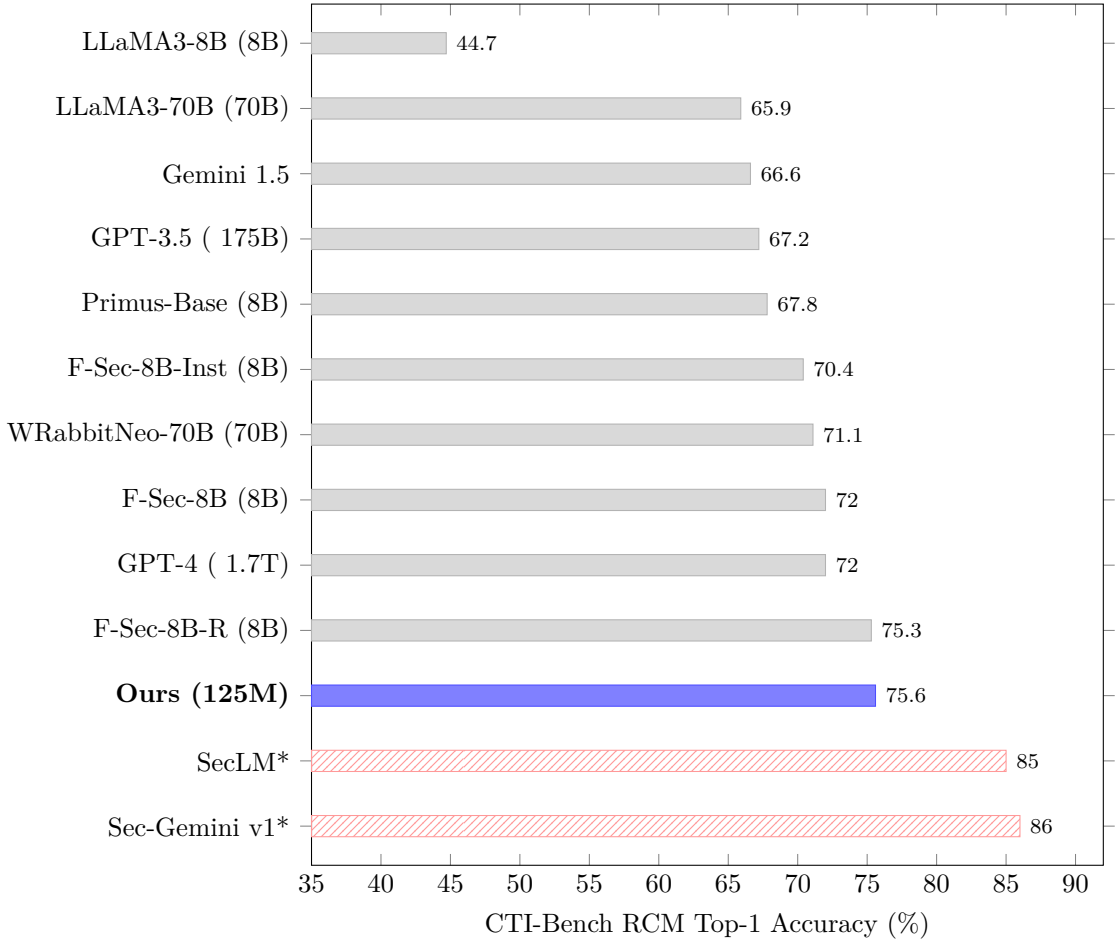

CTI-Bench \citep{alam2024ctibench} is an external benchmark with 2,000 CVEs (zero overlap with our training data). The RCM task evaluates CVE-to-CWE mapping using strict exact CWE ID match. It contains two splits: cti-rcm (1,000 CVEs from 2023--2024) and cti-rcm-2021 (1,000 CVEs from 2011--2021). Our training data spans 1999--2026, with 62.8\% from pre-2023 and 37.2\% from 2023+. While the specific CTI-Bench CVEs are excluded via decontamination, vulnerability patterns from the same period are present in training, so this is not a strict temporal split. Results are shown in Table~\ref{tab:ctibench} and Figure~\ref{fig:comparison}.

\begin{table}[h]
\centering
\caption{CTI-Bench RCM results (strict exact CWE match). 95\% Clopper-Pearson confidence intervals computed by us from reported accuracy and $n{=}1000$. $^*$Approximate scores estimated from published comparison charts; exact values not reported.}
\label{tab:ctibench}
\begin{tabular}{llllc}
\toprule
Model & Params & Type & cti-rcm (95\% CI) & Source \\
\midrule
Sec-Gemini v1 (Google)$^*$ & --- & closed & $\sim$86\% & \citet{google2025secgemini} \\
SecLM (Google)$^*$ & --- & closed & $\sim$85\% & \citet{google2025seclm} \\
\textbf{Ours} & \textbf{125M} & \textbf{open} & \textbf{75.6\%} (72.8--78.2) & --- \\
Foundation-Sec-8B-R (Cisco) & 8B & open & 75.3\% (72.5--77.9) & \citet{cisco2025reasoning} \\
GPT-4 & $\sim$1.7T & closed & 72.0\% & \citet{alam2024ctibench} \\
Foundation-Sec-8B (Cisco) & 8B & open & 72.0\% ($\pm$1.7\%) & \citet{cisco2025foundation} \\
WhiteRabbitNeo-V2-70B & 70B & open & 71.1\% & \citet{cisco2025foundation} \\
Foundation-Sec-8B-I (Cisco) & 8B & open & 70.4\% & \citet{cisco2025reasoning} \\
Llama-Primus-Base (Trend Micro) & 8B & open & 67.8\% & \citet{trendmicro2025} \\
GPT-3.5 & $\sim$175B & closed & 67.2\% & \citet{alam2024ctibench} \\
Gemini 1.5 & --- & closed & 66.6\% & \citet{alam2024ctibench} \\
LLaMA3-70B & 70B & open & 65.9\% & \citet{alam2024ctibench} \\
LLaMA3-8B & 8B & open & 44.7\% & \citet{alam2024ctibench} \\
\bottomrule
\end{tabular}
\end{table}

Our 125M-parameter model achieves 75.6\% (95\% CI: 72.8--78.2\%), overlapping with Cisco's Foundation-Sec-8B-Reasoning at 75.3\% (95\% CI: 72.5--77.9\%). The 0.3pp difference is \textbf{not statistically significant}---we claim competitive performance, not superiority. The key result is parameter efficiency: comparable task-specific accuracy with 64$\times$ fewer parameters, though we note that the Cisco models are general-purpose LLMs capable of many tasks beyond CWE classification.

\section{Analysis}

\subsection{CWE Hierarchy and the Granularity Gap}

A significant portion of our ``errors'' on CTI-Bench stem from a CWE granularity mismatch. Our model is trained on refined labels that prefer specific child CWEs (e.g., CWE-121 Stack Buffer Overflow), while CTI-Bench ground truth uses NVD labels that often assign generic parent CWEs (e.g., CWE-119 Buffer Overflow).

When we apply hierarchy-aware evaluation where parent$\leftrightarrow$child equivalences count as correct:

\begin{table}[h]
\centering
\caption{Hierarchy-aware evaluation on CTI-Bench. These scores are \textbf{not directly comparable} to strict scores from other models and are presented as supplementary analysis.}
\label{tab:hierarchy}
\begin{tabular}{lccc}
\toprule
Benchmark & Strict & Hierarchy-aware & Rescued \\
\midrule
cti-rcm (2024) & 75.6\% & 86.5\% (+10.9pp) & 109 samples \\
cti-rcm-2021 & 71.8\% & 82.8\% (+11.0pp) & 110 samples \\
\bottomrule
\end{tabular}
\end{table}

For example, predicting CWE-122 (Heap Buffer Overflow) when the ground truth is CWE-787 (Out-of-bounds Write) is technically a more specific and arguably more useful prediction. We emphasize that hierarchy-aware scores are \textbf{not directly comparable} to strict scores from other models.

\subsection{Internal vs External Evaluation Gap}

Our model achieves 87.4\% on the internal test set but 75.6\% (strict) on CTI-Bench. This 11.8pp gap is explained by:

\begin{enumerate}[nosep]
    \item \textbf{Agreement filtering} (primary factor): Our test set contains only samples where NVD and Sonnet agree---by construction, these are ``easier'' samples where two independent labelers concur. CTI-Bench has no such filter, and includes the ambiguous cases our test set excludes.
    \item \textbf{CWE granularity mismatch} (10.9pp): Hierarchy-aware evaluation closes nearly the entire gap (86.5\% vs 87.4\%).
    \item \textbf{Label noise in ground truth}: CTI-Bench uses raw NVD labels, which disagree with independent AI relabeling 26.9\% of the time on exact match (15.5\% when allowing hierarchy-aware equivalences).
\end{enumerate}

\subsection{Ablation: Two-Phase Training}

We validated the two-phase approach through systematic experiments on the validation set:

\begin{table}[h]
\centering
\caption{Ablation study: effect of two-phase training.}
\label{tab:ablation}
\begin{tabular}{lc}
\toprule
Configuration & Val Top-1 \\
\midrule
Standard fine-tuning (no freezing) & 87.7\% \\
Two-phase: freeze-6 + full & 87.9\% \\
\textbf{Two-phase: freeze-8 + full} & \textbf{88.0\%} \\
Two-phase: freeze-8 + full (more epochs) & 87.9\% \\
\bottomrule
\end{tabular}
\end{table}

Freezing 8/12 layers in phase~1 provides the best trade-off, allowing the classifier head to stabilize before full fine-tuning.

\subsection{Per-Class Performance on CTI-Bench}

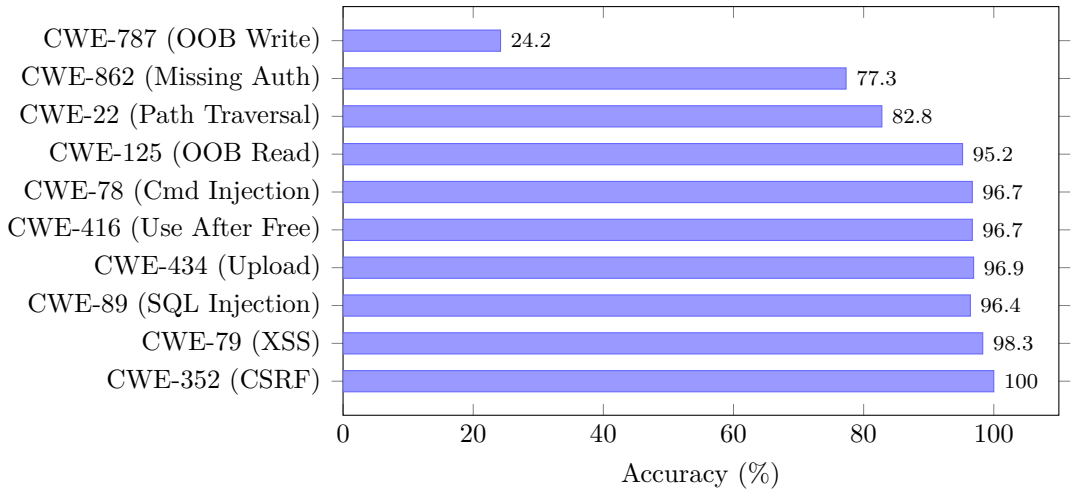
\begin{figure}[h]
\centering
\begin{tikzpicture}
\begin{axis}[
    xbar,
    width=11cm,
    height=7cm,
    xlabel={Accuracy (\%)},
    xmin=0,
    xmax=110,
    ytick=data,
    yticklabels={
        CWE-787 (OOB Write),
        CWE-862 (Missing Auth),
        CWE-22 (Path Traversal),
        CWE-125 (OOB Read),
        CWE-78 (Cmd Injection),
        CWE-416 (Use After Free),
        CWE-434 (Upload),
        CWE-89 (SQL Injection),
        CWE-79 (XSS),
        CWE-352 (CSRF),
    },
    y dir=reverse,
    bar width=8pt,
    nodes near coords,
    nodes near coords align={horizontal},
    every node near coord/.append style={font=\scriptsize},
    tick label style={font=\small},
    label style={font=\small},
]
\addplot[fill=blue!40, draw=blue!60] coordinates {
    (24.2, 0)
    (77.3, 1)
    (82.8, 2)
    (95.2, 3)
    (96.7, 4)
    (96.7, 5)
    (96.9, 6)
    (96.4, 7)
    (98.3, 8)
    (100.0, 9)
};
\end{axis}
\end{tikzpicture}
\caption{Per-CWE accuracy on CTI-Bench cti-rcm (top 10 classes by support). CWE-787 is an outlier due to the hierarchy issue---the model predicts specific children (CWE-121, CWE-122) that map to CWE-787.}
\label{fig:perclass}
\end{figure}

Performance is excellent ($>$95\%) on well-defined CWEs with distinctive vocabulary. The outlier is CWE-787 (24.2\%), which is almost entirely due to the hierarchy issue---the model correctly predicts specific children (CWE-121, CWE-122) that map to CWE-787 in NVD's labeling.

\subsection{Manual Validation of Label Quality}
\label{sec:manual}

To address the circularity concern (``why trust Sonnet over NVD?''), we manually reviewed 100 randomly sampled disagreement cases from the full population of 74,677 NVD-Sonnet disagreements. For each case, the first author examined the CVE description, both CWE labels, and Sonnet's reasoning to determine which label better matches the described vulnerability.

\begin{table}[h]
\centering
\caption{Manual review of 100 random NVD-Sonnet disagreement cases.}
\label{tab:manual}
\begin{tabular}{lc}
\toprule
Verdict & Count \\
\midrule
Sonnet clearly more accurate & 72 \\
Both acceptable (hierarchy/sibling CWEs) & 16 \\
Ambiguous (both generic or unclear) & 9 \\
NVD clearly more accurate & 3 \\
\bottomrule
\end{tabular}
\end{table}

In 72\% of disagreements, Sonnet assigned a more specific or more accurate CWE than NVD. The most common patterns: NVD using deprecated/generic catch-all CWEs (CWE-20, CWE-119, CWE-264, CWE-399) where Sonnet assigned the correct specific CWE; NVD assigning the wrong CWE category entirely (e.g., CWE-79 XSS for a CSRF vulnerability); and Sonnet selecting the precise child CWE (e.g., CWE-121 Stack Buffer Overflow vs generic CWE-119). NVD was clearly more accurate in only 3 cases---two involving CWEs outside Sonnet's 205-class vocabulary, and one where the description explicitly ruled out the Sonnet-predicted category. This provides evidence that the Sonnet-refined labels used for training are higher quality than raw NVD labels, though we note the review was performed by a single non-specialist annotator.

\section{Related Work}

\textbf{CVE-to-CWE classification.}\quad V2W-BERT \citep{na2021v2wbert} reports 94--97\% accuracy on NVD data with 124 CWE classes, but uses random train/test splits on historical data where similar CVEs (e.g., same product, same vulnerability pattern) frequently appear in both splits---a form of data leakage that inflates reported accuracy. Our evaluation uses temporal decontamination (CTI-Bench) and agreement-filtered labels, which are strictly harder settings. \citet{cvedrill2024} report 82\% vs ChatGPT's 56\% on a custom evaluation set with a different class taxonomy. CTI-Bench \citep{alam2024ctibench} provides the first standardized benchmark.

\textbf{Cybersecurity language models.}\quad SecureBERT \citep{aghaei2023securebert}, CyBERT \citep{ranade2021cyberta}, and Foundation-Sec-8B \citep{cisco2025foundation} are pretrained on cybersecurity corpora. Our approach uses a general-purpose pretrained model with task-specific fine-tuning and high-quality labels, achieving competitive results. A natural question is whether domain-specific encoders (e.g., SecureBERT, DeBERTa-v3) would perform better as a base model; we leave this comparison to future work.

\textbf{LLM-based approaches.}\quad Google's Sec-Gemini~v1 \citep{google2025secgemini} reports the highest CTI-Bench RCM performance, outperforming other models by at least 10.5\% (approximately 83--86\% based on published charts), but is a closed proprietary system likely augmented with retrieval over CWE taxonomy descriptions. Our work shows that a small fine-tuned encoder can achieve competitive open-weight results.

\section{Limitations}

\begin{itemize}[nosep]
    \item \textbf{Single-label CWE}: Each CVE is assigned exactly one CWE, though some vulnerabilities involve multiple weakness types.
    \item \textbf{205 CWE classes}: Covers the most common CWEs in NVD. Rare CWEs not in the training set will be mapped to the closest known class.
    \item \textbf{English only}: Trained on English NVD descriptions.
    \item \textbf{Description-based}: Uses only text descriptions, not CVSS scores, CPE, or other structured metadata.
    \item \textbf{Hierarchy-aware evaluation}: Our supplementary hierarchy-aware metric is not standardized and may not be directly comparable to other work.
    \item \textbf{Label circularity}: We use Sonnet labels for training and Sonnet-NVD agreement for evaluation. While the external CTI-Bench evaluation is fully independent of Sonnet, our internal test set metrics may partially reflect Sonnet's biases. Our manual validation (Section~\ref{sec:manual}) mitigates but does not fully eliminate this concern, as the reviewer is not a CWE taxonomy expert.
    \item \textbf{No base model comparison}: We did not evaluate alternative encoders (DeBERTa-v3, SecureBERT) fine-tuned on the same data, which would isolate the contribution of label quality vs.\ architecture choice.
\end{itemize}

\section{Conclusion}

We demonstrate that a task-specific fine-tuned encoder (125M parameters) can achieve competitive open-weight performance on CVE-to-CWE classification, matching Cisco's 8B-parameter general-purpose Foundation-Sec-8B-Reasoning on CTI-Bench (75.6\% vs 75.3\%, difference not statistically significant) with 64$\times$ fewer parameters. The key ingredients are high-quality AI-refined training labels and two-phase fine-tuning.

We identify CWE hierarchy granularity mismatch as a significant confound in current benchmarks and argue that models predicting specific child CWEs should not be penalized for disagreeing with generic parent labels.

All resources are publicly available:
\begin{itemize}[nosep]
    \item Dataset: \url{https://huggingface.co/datasets/xamxte/cve-to-cwe}
    \item Model: \url{https://huggingface.co/xamxte/cwe-classifier-roberta-base}
\end{itemize}

\bibliographystyle{plainnat}

\begin{thebibliography}{10}

\bibitem[Aghaei et~al.(2023a)]{aghaei2023securebert}
E.~Aghaei, X.~Niu, W.~Shadid, and E.~Al-Shaer.
\newblock SecureBERT: A Domain-Specific Language Model for Cybersecurity.
\newblock In \emph{Proc.\ 18th EAI International Conference on Security and Privacy in Communication Networks (SecureComm)}, LNICST vol.~462, pp.~39--56. Springer, 2023a.

\bibitem[Alam et~al.(2024)]{alam2024ctibench}
M.~T. Alam, D.~Bhusal, L.~Nguyen, and N.~Rastogi.
\newblock CTIBench: A Benchmark for Evaluating LLMs in Cyber Threat Intelligence.
\newblock \emph{Advances in Neural Information Processing Systems}, 37:50805--50825, 2024.

\bibitem[Aghaei et~al.(2023b)]{cvedrill2024}
E.~Aghaei, E.~Al-Shaer, W.~Shadid, and X.~Niu.
\newblock Automated CVE Analysis for Threat Prioritization and Impact Prediction.
\newblock arXiv:2309.03040, 2023b.

\bibitem[Kassianik et~al.(2025)]{cisco2025foundation}
P.~Kassianik, B.~Saglam, A.~Chen, B.~Nelson, et~al.
\newblock Llama-3.1-FoundationAI-SecurityLLM-Base-8B Technical Report.
\newblock arXiv:2504.21039, 2025.

\bibitem[Yang et~al.(2026)]{cisco2025reasoning}
Z.~Yang, E.~Li, J.~He, A.~Priyanshu, et~al.
\newblock Llama-3.1-FoundationAI-SecurityLLM-Reasoning-8B Technical Report.
\newblock arXiv:2601.21051, 2026.

\bibitem[Google(2025a)]{google2025secgemini}
Google.
\newblock Sec-Gemini v1.
\newblock Google Security Blog, April 2025.
\newblock \url{https://security.googleblog.com/2025/04/google-launches-sec-gemini-v1-new.html}.

\bibitem[Google(2025b)]{google2025seclm}
Google.
\newblock Fueling AI Innovation in SecOps Products: The SecLM Platform.
\newblock Google Cloud Community Blog, 2025.

\bibitem[Liu et~al.(2019)]{liu2019roberta}
Y.~Liu, M.~Ott, N.~Goyal, J.~Du, M.~Joshi, D.~Chen, O.~Levy, M.~Lewis, L.~Zettlemoyer, and V.~Stoyanov.
\newblock RoBERTa: A Robustly Optimized BERT Pretraining Approach.
\newblock arXiv:1907.11692, 2019.

\bibitem[Ranade et~al.(2021)]{ranade2021cyberta}
P.~Ranade, A.~Piplai, A.~Joshi, and T.~Finin.
\newblock CyBERT: Contextualized Embeddings for the Cybersecurity Domain.
\newblock In \emph{Proc.\ 2021 IEEE International Conference on Big Data}, pp.~3334--3342, 2021.

\bibitem[Das et~al.(2021)]{na2021v2wbert}
S.~S. Das, E.~Serra, M.~Halappanavar, A.~Pothen, and E.~Al-Shaer.
\newblock V2W-BERT: A Framework for Effective Hierarchical Multiclass Classification of Software Vulnerabilities.
\newblock In \emph{Proc.\ 2021 IEEE 8th International Conference on Data Science and Advanced Analytics (DSAA)}, pp.~1--12, 2021.

\bibitem[Yu et~al.(2025)]{trendmicro2025}
Y.-C. Yu, T.-H. Chiang, C.-W. Tsai, C.-M. Huang, and W.-K. Tsao.
\newblock Primus: A Pioneering Collection of Open-Source Datasets for Cybersecurity LLM Training.
\newblock In \emph{Proc.\ 2025 Conference on Empirical Methods in Natural Language Processing (EMNLP)}, pp.~10402--10424, 2025.

\end{thebibliography}

\end{document}